\documentclass[prb,twocolumn,showpacs,floatfix,amsmath]{revtex4-1}

\usepackage{color}
\usepackage{dsfont}
\usepackage[normalem]{ulem}
\usepackage{dcolumn}

\usepackage{graphicx}
\def\be{\begin{equation}}
\def\ee{\end{equation}}
\def\beq{\begin{eqnarray}}
\def\eeq{\end{eqnarray}}

\newcommand{\ket}[1]{\ensuremath{\lvert #1 \rangle}}
\newcommand{\bra}[1]{\left\langle#1\right|}

\begin{document}
\renewcommand{\familydefault}{\sfdefault}
\renewcommand{\sfdefault}{cmbr}
\title{Giant perpendicular magnetic anisotropy energies in CoPt thin films: Impact of 
reduced dimensionality and imperfections}
\author{Samy Brahimi$^1$}\email{samy86brahimi@gmail.com}
\author{Hamid Bouzar$^1$}
\author{Samir Lounis$^2$}
\affiliation{$^1$Laboratoire de Physique et Chimie Quantique, Universit\'e Mouloud Mammeri, Tizi-Ouzou, 15000 Tizi-Ouzou, Algeria}
\affiliation{$^2$ Peter Gr\"unberg Institut and Institute for Advance Simulation, Forschungszentrum J\"ulich, 52425 J\"ulich \& JARA, Germany}

\begin{abstract}
The impact of reduced dimensionality on the magnetic properties of the tetragonal 
{\bf{L1}$_{0}$} CoPt alloy is investigated from ab-initio 
considering several kinds of surface defects.  
By exploring the dependence of the magnetocrystalline anisotropy energy (MAE) on the 
thickness of CoPt thin films, we demonstrate the crucial role of the chemical nature of the 
surface. For instance, Pt-terminated thin films exhibit huge MAEs which can be 1000\% 
larger than those of Co-terminated films. Besides 
the perfect thin films, we scrutinize the effect of defective surfaces such as stacking 
faults or anti-sites on the surface layers. Both types of defects reduce considerably 
the MAE with respect to the one obtained for Pt-terminated thin films. A detailed analysis of the 
electronic structure of the thin films is provided with a careful comparison to the CoPt bulk case. 
The behavior of the MAEs is then related to the location of the different virtual bound states utilizing 
second order perturbation theory.
\end{abstract}      

\maketitle
\date{\today}

\section{Introduction}
\begin{figure}[htpb]
\centering
\includegraphics[width=\columnwidth]{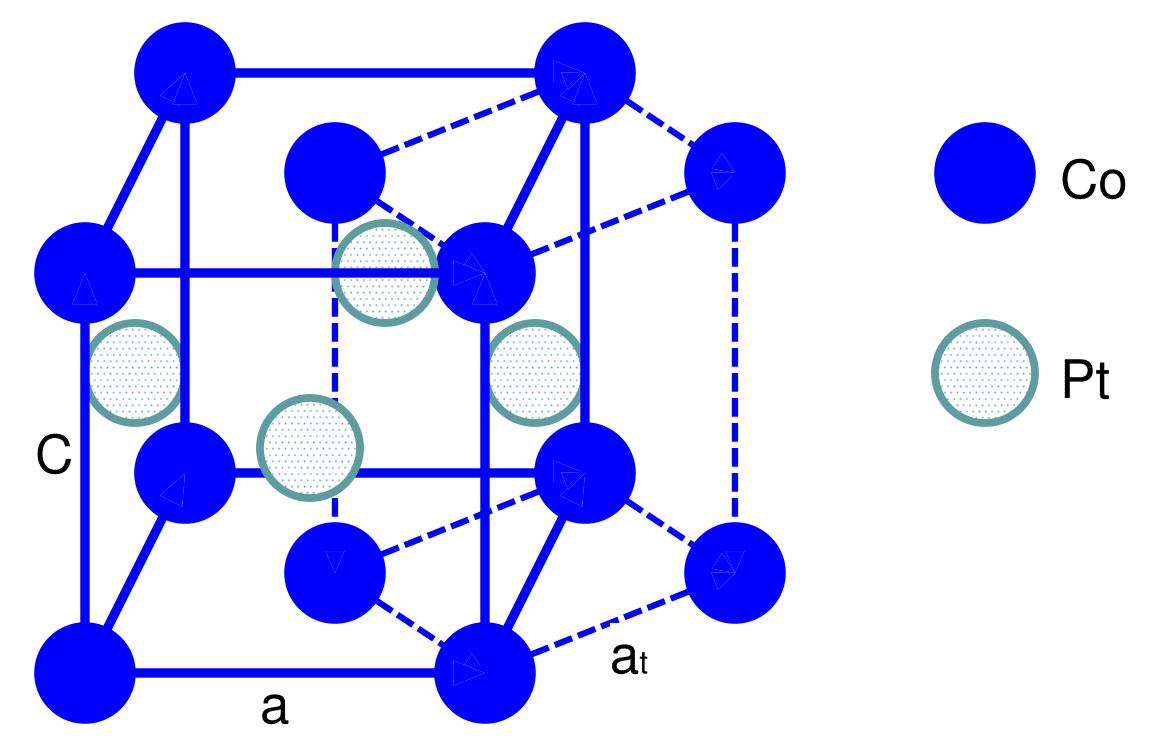} 
\caption{The conventional cell of the CoPt {\bf{L1}$_{0}$} alloy. The primitive cell is also sketched using dashed lines.}
\label{cell_bulk}
\end{figure}

The magnetocrystalline anisotropy energy (MAE) is at the heart of magnetic properties of materials. 
It is of crucial importance from the 
fundamental or technological point of views since it provides an energy scale for the stability of
 magnetic domains where for example magnetic information is stored. 
When the MAE is large and favors an out-of-plane orientation of the magnetic moments, perpendicular magnetic recording or 
magneto-optical recording is possible (see e.g. Refs\cite{Coffey1995,Sellmyer2005}). 
CoPt binary bulk alloy in the {\bf{L1}$_{0}$} structure (see Fig.\ref{cell_bulk}) is by now a 
classical example of a material exhibiting a large perpendicular MAE, around 1 meV ~\cite{Maykov1989,Ye1990,Weller1999,Yang2004,Perez2005}. 
There has been a 
tremendous number of studies related to the magnetic properties 
of this alloy 
in its bulk phase, as nanoparticles or in nanostructures combining Co and Pt 
(see e.g. Refs.~\cite{Coffey1995,Szunyogh1999,Uba2002,Park2004,Bartolome2007,Tournus2008,Entel2009,Alloyeau2009,Blanc2011,Ouazi2012,
Karoui2013,Zemen2014,
Figueroa2014,Hu2014}).
%Figure 1

A large amount of work has been devoted to unveil the origin of the large perpendicular MAE in binary bulk alloys, 
see e.g. Refs\cite{Daalderop1991,Sakuma1994,Solovyev1995,Oppeneer1998,Razee1999,Galanakis2000,Ravindran2001,Schick2003}. 
The interplay of the tetragonality of the alloy, band filling, hybridization between the constituents affect 
certainly the magnitude of the MAE. For instance, tetragonality leads to the lifting of the degeneracy of 
the electrons by the tetragonal crystal field and produces thereby an additional contribution to the MAE. 
Thus, and as expected from perturbation theory, the MAE becomes 
proportional to $\xi^2$ instead of $\xi^4$ as found for cubic symmetry, where $\xi$ is the 
spin-orbit coupling constant. Indeed, in cubic bulk systems, the high symmetry allows only for a fourth-order anisotropy 
constant, and thus  they are characterized by a small MAE. Razee et al.\cite{Razee1999} argued however 
that the tetragonal distortion of CoPt, given by the axial ration c/a = 0.98, 
 contributes by only 15\% of the MAE. It was then concluded that the compositional order of the alloy is an important ingredient for a large MAE.

Sakuma\cite{Sakuma1994} shows that by changing the axial ratio (c/a) defining the 
tetragonality of CoPt and FePt alloys, the MAE first smoothly decreases by increasing c/a 
till reaching a minimum at $\sim$0.8 before a smooth increase in magnitude. Interestingly, except a small window of axial ratios ($0.7 < c/a < 0.9$), 
the MAE favors an out-of-plane orientation of the magnetic 
moments. The tetragonalization is then thought to provide an effect similar to the band filling\cite{Sakuma1994,Daalderop1991}.

In the context of thin films, Zhang et al.\cite{Zhang2009} demonstrated with ab-initio simulations that for CoPt films terminated by Co layers, 
a thickness of at least 9 monolayers exhibit a rather converged MAE, with a bulk contribution of 1.36 meV favoring 
a perpendicular orientation of the magnetic moments and 
a counter-acting 
surface contribution of -0.76 meV favoring, interestingly, an in-plane orientation of the moments. Their 
interest in CoPt was motivated by the experimental 
demonstration of coercivity manipulation of {\bf{L1}$_{0}$} FePt and FePd thin films\cite{Weisheit2007} 
by external electric field. Their ab-initio 
simulations predicted a higher sensitivity of CoPt to electric field than that of FePt films. 
Pustogowa et al.\cite{Pustogowa1999} investigated from first-principles  several components made of 
Co and Pt deposited on Pt(100) and Pt(111) surfaces. They found 
that ordered superstructures of (CoPt)$_n$ deposited on both mentioned substrates are characterized 
by a perpendicular MAE, which is heavily affected by chemical 
disorder in line with the analysis of Razee et al.\cite{Razee1999}.
%Figure 2
\begin{figure*}[htpb]
\centering
\includegraphics[width=2\columnwidth]{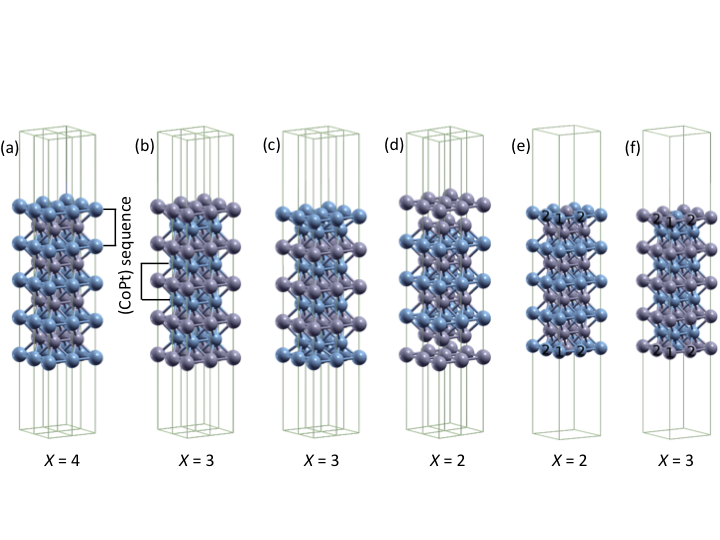} 
\caption{Supercells used for the simulation of the (001) CoPt thin films where the blue and 
magenta spheres correspond respectively 
to the Co and Pt atoms: (a) Pure Co surface, (b) pure Pt surface, (c) Co stacking fault, (d) Pt stacking fault, 
{(e) Pt anti-site and (f) Co anti-site}. In the latter two cases, 
numbers 1 and 2 refer to atoms with different magnetic moments. 
{For each case, the number of (CoPt) sequences, $X$, is given.}}
\label{cell_surf}
\end{figure*}

The goal of this manuscript is to present a systematic ab-initio investigation on the effect 
of reduced dimensionality on the magnetic properties of CoPt(001) 
films with a focus on their MAE and by addressing the impact of the termination type of the 
films. Contrary to previous investigations\cite{Zhang2009},  
we consider not only 
Co-terminated films but also Pt-terminated films and several types of surface defects (see Fig.\ref{cell_surf}).  
For instance, we found that decreasing the thickness of the films leads to a sign change of the surface MAE. 
 Pt covered thin films can boost the total perpendicular MAE by a large amount stabilizing, thereby, 
more strongly the out-of-plane orientation of the moments. 
Molecular dynamics simulations demonstrated 
the likeliness of having Pt on the surface of CoPt alloy\cite{Ersen2008,Liu2016} and thus the pertinence of our predictions. 
After a careful study of different defective terminations types (stacking faults, anti-site defects), 
we provide the ingredient to increase the MAE 
of the thin films. If we label the Co and Pt layers by respectively A and B, the perfect 
stacking along the [001] direction is given for example by ABABAB for 6 layers. 
Possible stacking faults, which are planar defects, could be the sequence ABABAA (see Figs.\ref{cell_surf}(c-d)). 
Anti-site defects on the surface means that instead of having at the surface a pure layer A, or layer 
B, we have an alloy, for example, made of A and B. In our work, we considered an alloy 
of the type A$_{0.25}$B$_{0.75}$ in the  surface layer instead of the 
perfect B layer of our example (see Figs.\ref{cell_surf}(e-f)). 

\section{Method}
We simulate the thin films by adopting the slab approach with periodic boundary 
conditions in two directions while the periodic images in the third direction are 
separated by a sufficient amount of vacuum (15 \AA~) to avoid  interaction between 
neighboring supercells. We have chosen to use symmetrical calculation cells 
with an odd number of planes to avoid the pulay stress. Some representative slabs are shown in Fig.\ref{cell_surf}. 
Here it can be observed that for equiatomic {\bf{L1}$_{0}$} type of alloys two different 
surfaces exist when the slabs are stacked along the {[001]} direction. In the perfect cases, the surface termination can be made of either purely Co atoms or Pt atoms. 
The self-consistent calculations are carried out with the Vienna ab initio simulation package (VASP) 
using a plane wave basis and the projector augmented wave (PAW) approach~\cite{PAW1,PAW2}. 
The exchange-correlation potential is used in the functional from of Perdew, Burke and Ernzerhof (PBE)~\cite{PBE1,PBE2}. 
The cut-off energies for the plane waves is 478 eV. The integration over the Brillouin zone was based on finite temperature smearing (Methfessel-Paxton method) for the thin films 
while for the bulk case the tetrahedron method with Bl\"ochl corrections has been used. 
The k-points grids are \(14\times14\times11\) for the bulk calculation, and \(14\times14\times1\) for the {(001)} surface calculations. The energy 
convergence criterion is set to \(10^{-8}\) eV while the geometrical atomic relaxations for the surfaces calculations were stopped when the forces were less than 0.01 eV/\AA. 
The MAE is extracted from the difference between the total energies of the two configurations: out-of-plane versus in-plane orientations of the 
magnetic moments. A positive value indicates a preference for the 
out-of-plane orientation of the magnetic moments. 
\begin{figure}[htpb]
\centering
\includegraphics[width=\columnwidth]{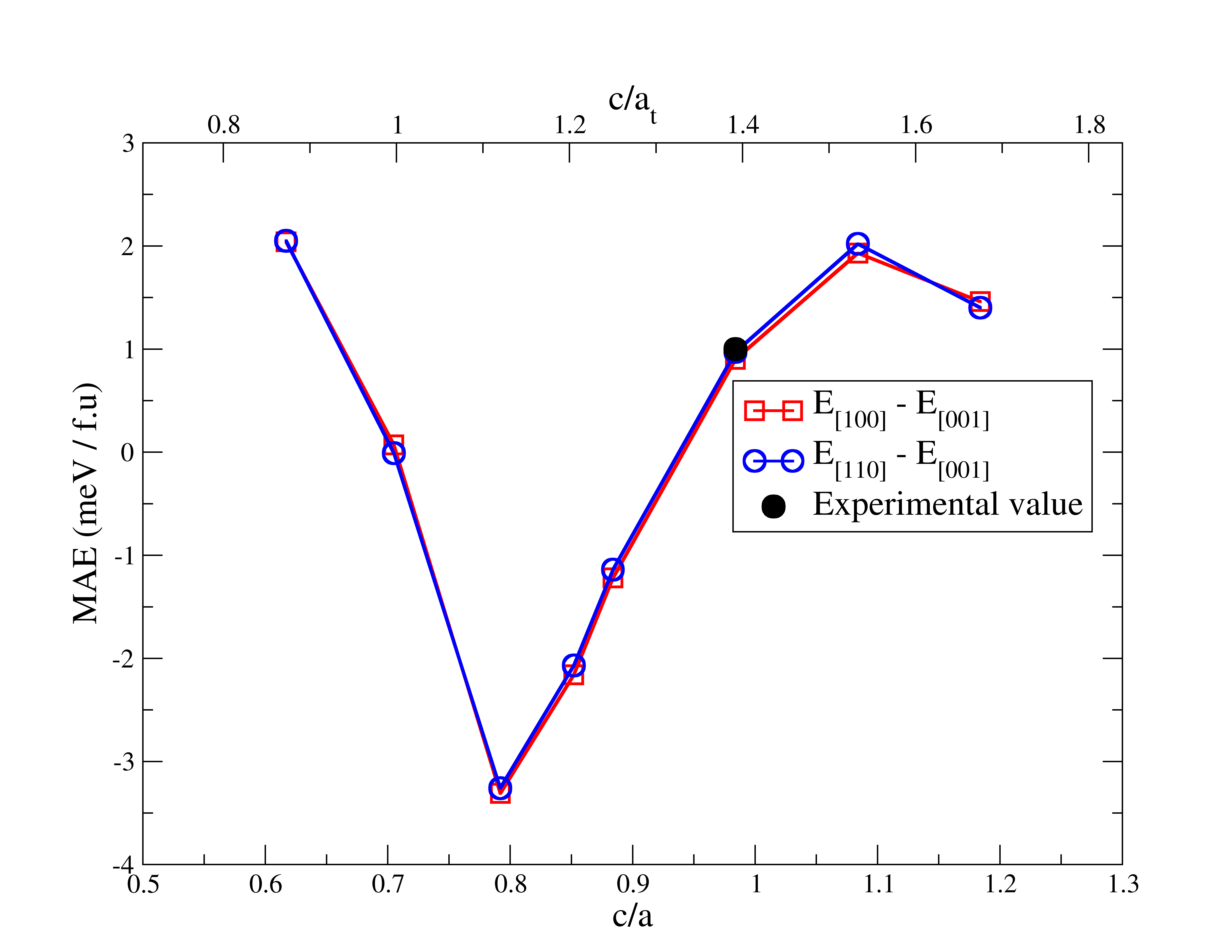} 
\caption{MAE of the bulk $L1_0$ CoPt alloy as function of the axial ratio c/a under constant volume. Two possible in-plane 
orientation of the magnetic moments are considered, [100] and [110], but the obtained MAE are very similar. The 
closed circle represents the experimental value\cite{Eurin1969}, which is well reproduced by our simulations. Other experimental values can be 50\% larger, 
see e.g. Ref.~\cite{Grange2000}.}
\label{MAE_bulk}
\end{figure}

\section{Results} 

\subsection{Bulk CoPt}
To start our investigations, we revisited the bulk alloy phase by finding the tetragonal lattice structure minimizing its energy. 
The optimal value of c/a ratio is equal to 0.984 with a lattice parameter value a of 3.80 \AA, in good 
agreement with values available in the literature (see e.g. \cite{Karoui2013,Pearson1964,Galanakis2000,Ravindran2001}). 
The calculated magnetic moments (\(M_{\mathrm{Co}}\) = 1.91 \(\mu_{B}\), 
\(M_{\mathrm{Pt}}\) = 0.40 \(\mu_{B}\)) are also in line with previous works~\cite{Cadeville1987,Dannenberg2009,Dupuis2015,Galanakis2000,Ravindran2001}. 
We also note the well-known emergence of an induced moment in Pt, which is due to 
the hybridization of its  5d orbitals with the exchange splitted 3d orbitals of Co. 
In order to calculate the MAE, we considered two possible directions for the in-plane moments orientations, [110] and [100], 
and we found a negligible difference in 
the obtained perpendicular MAE.
For the optimized structure, the MAE reaches a value 
of 0.91 meV when the in-plane orientation of moments is along [100] and 0.97 meV for [110] as an in-plane orientation of 
the moments. Both values are close to the experimental values that are given around 1 meV~\cite{Eurin1969,Grange2000}.

In Fig.\ref{MAE_bulk}, we plot the bulk MAE as function of the ratio c/a under the constant unit cell volume in a similar 
fashion then that of Sakuma~\cite{Sakuma1994}. The obtained curve agrees well with the one published in the latter article. 
For ratios between 0.6 and 1.2, the MAE experiences one minimum 
and two maxima. The largest in-plane MAE is found for a ratio of 0.8. As expected, if c/a = $1/\sqrt{2}$, i.e. c/a$_t$ = 1, the MAE drops to 
zero since this corresponds to a primitive cell of the cubic B2 structure. As discussed by Sakuma, upon tetragonalization, 
the electronic states of the atoms are shifted and the band filling changes, which affect the magnitude and sign of the MAE. 
\begin{figure*}[htpb]
\centering
 \begin{tabular}{c}
\includegraphics[width=1.1\columnwidth]{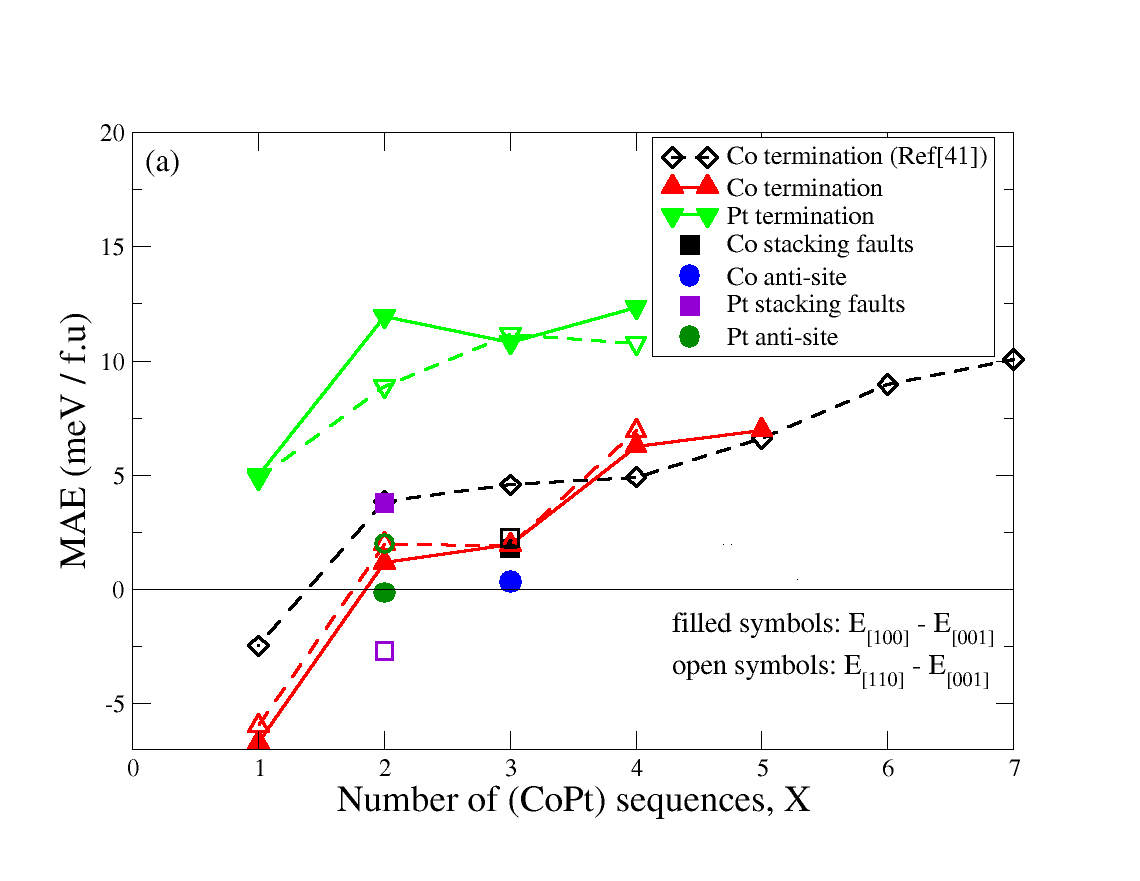}\hspace{-1.cm}
\includegraphics[width=1.1\columnwidth]{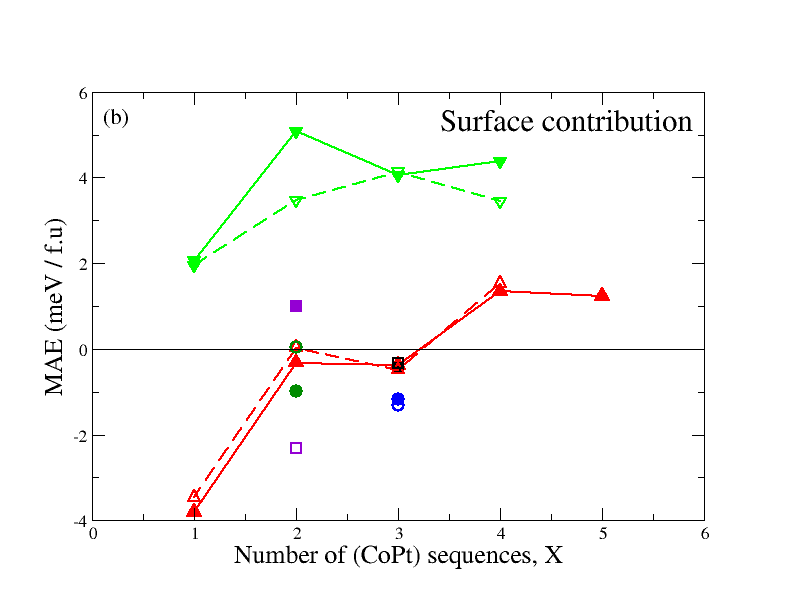} 
\end{tabular}
\caption{{MAE of CoPt thin films as function of $X$, the number of (CoPt) sequences. 
In contrast to (a), where the MAE of total thin films is plotted, in (b) the surface contribution is depicted.
Several cases are considered: Co-terminated 
thin films (red triangles), Pt-terminated thin films (green triangles), stacking faults defects (Co with a black square and 
Pt with a magenta square), anti-sites (Co with a blue circle and Pt with a green circle). The diamonds represent 
the data of Zhang et al.\cite{Zhang2009} obtained for Co-terminated thin films considering the MAE with respect to the direction 
[110]. For completeness, we consider both type of possible in-plane orientation of the moments, along the [110] shown with open symbols 
and along the [100] direction with filled symbols.}}
\label{MAE_total}
\end{figure*}
%Table 2. 
\begin{table*}
\begin{tabular}{|p{1.2cm}|p{1.2cm}|p{1.2cm}|}
\hline
	Layer & Atom & M (\(\mu_{B}\))\\\hline
	S & Co & 2.00\\\hline
	S-1 & Pt & 0.40\\\hline
	S-2 & Co & 1.90\\\hline
	S-3 & Pt & 0.40\\\hline
	Center & Co & 1.90\\\hline
\end{tabular}
\begin{tabular}{|p{1.5cm}|}
\hline
	\rule[0cm]{0mm}{-0.3cm}\(\delta\)\emph{d}/\(d_{0}(\%)\) \\\hline
	\rule[0cm]{0mm}{-0.3cm}-4.30  \\\hline
	\rule[0cm]{0mm}{-0.3cm} 1.40 \\\hline
	\rule[0cm]{0mm}{-0.3cm}-0.80 \\\hline
	\rule[0cm]{0mm}{-0.3cm}-0.05 \\\hline
\end{tabular}\hspace{1cm}
\begin{tabular}{|p{1.2cm}|p{1.2cm}|p{1.2cm}|}
\hline
	Layer & Atom & M (\(\mu_{B}\))\\\hline
	S & Pt & 0.42\\\hline
	S-1 & Co & 1.99\\\hline
	S-2 & Pt & 0.40\\\hline
	S-3 & Co & 1.90\\\hline
	Center & Pt & 0.40\\\hline
\end{tabular}
\begin{tabular}{|p{1.5cm}|}
\hline
	\rule[0cm]{0mm}{-0.3cm}\(\delta\)\emph{d}/\(d_{0}\) (\%)\\\hline
	\rule[0cm]{0mm}{-0.3cm}-5.60  \\\hline
	\rule[0cm]{0mm}{-0.3cm} 1.60 \\\hline
	\rule[0cm]{0mm}{-0.3cm}-0.50 \\\hline
	\rule[0cm]{0mm}{-0.3cm}-0.05 \\\hline
\end{tabular}
\\
\caption{Magnetic profile and geometrical relaxations of 
9 layers think thin film of CoPt terminated by Co (a) and by Pt (b). S labels the outermost surface layer and  
\(\delta\)\emph{d} = \emph{d} - \(d_{0}\) describes the changes of the interlayer distance $d$ with respect to that of the bulk $d_0$ that is 
equal to 1.871 \AA.}
\label{table_perfect}
\end{table*}

\subsection{Perfect surfaces}
As mentioned earlier, we considered different film thicknesses. We start analyzing the results obtained for perfect Co-terminated thin films. 
A representative of one of the simulated films is shown in Fig.~\ref{cell_surf}(a). 
We plot the MAE versus the film thickness in a fashion similar to 
that of Zhang et al.~\cite{Zhang2009}, i.e. considering along the x-axis the number of (CoPt) sequences, $X$ (see Fig.~\ref{MAE_total}(a)). { A single CoPt sequence is shown in Fig.~\ref{cell_surf}.}
Thus, for the Co-terminated 9 layers-thick thin film, $X = 4$. We notice that the  MAEs of the Co-terminated thin 
films are characterized by an oscillating behavior, which 
is induced by confinement effects. Indeed and as indicated by Zhang et al.~\cite{Zhang2009},  quantum well states can occur because of confinement, 
which can impact on the electronic structure of the thin films and thus 
on the related MAE. For $X=1$, the moments prefer interestingly 
an in-plane orientation contrary to thicker films.

As done for the bulk, two possible in-plane orientations of the magnetic moments are assumed: [100] and [110]. They are quasi-equivalent with 
a slight preference for the [100] direction. Thus, in the rest of our analysis we focus on the latter direction.
 
The MAE can be decomposed, as usually done, into a bulk contribution, $K_b$, and a surface contribution, $K_s$:

\begin{equation}
\mathrm{MAE} = X \times K_{b} + 2 \times K_{s}.
\end{equation}

 While this decomposition is  reasonable for thick films, it is questionable for the very thin films considered in our work. The 
value of $K_s$ extracted from the previous formula is then meant to indicate the impact of low dimensionality on the total MAE. 
One sees in Fig.~\ref{MAE_total}(b), that for $X < 4$ the surface interestingly contributes 
with a negative value to the total MAE and counteracts the ``bulk'' contribution.
%Figure 3
\begin{figure*}[htpb]
\centering
\includegraphics[width=2\columnwidth]{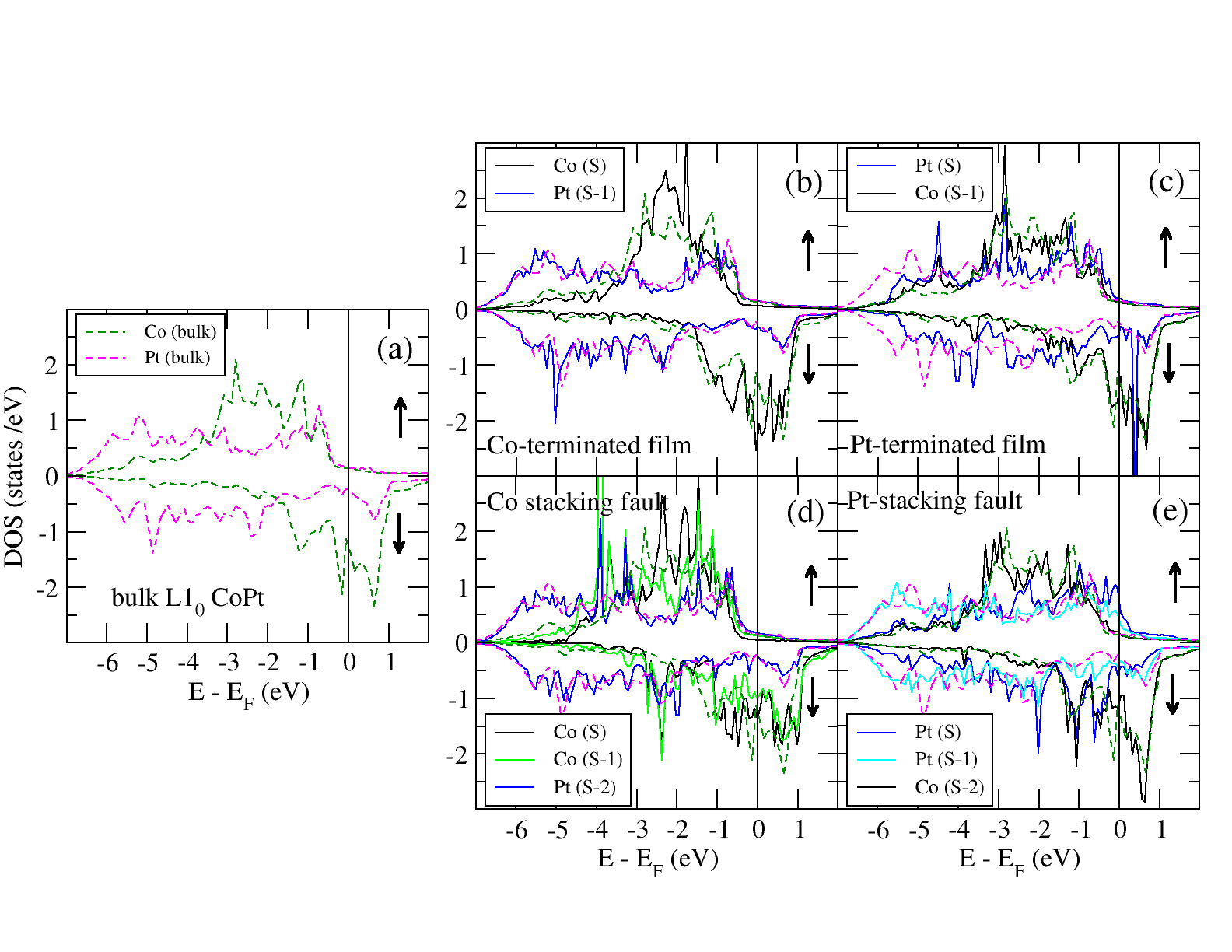}
\caption{Spin-dependent atom-projected electronic densities of states of CoPt {\bf{L1}$_{0}$} in the bulk phase (a), for 
Co (b) and Pt (c) terminated thin films, and films with Co (d) and Pt (e) stacking faults. S labels the top surface layer of the thin films.}
\label{DOS_total}
\end{figure*}

In order to provide a common reference to compare all kind of considered film terminations, 
we extend the concept used for the Co-terminated films and 
plot the MAEs as function of the (CoPt) sequences. A 9-layers thick Pt-terminated film is then characterized 
by $X=3$. In Fig.\ref{MAE_total}, it would then be compared to 
the Co-terminated film containing 7 atomic layers. This comparison indicates therefore the impact 
of Pt coverage of the Co-terminated film. 
A general observation deduced from Fig.~\ref{MAE_total} is the large value of the MAEs after Pt 
coverage, in some cases 1000\% larger than the one obtained for the Co-terminated films. Interestingly, 
the surface contribution to the total MAE favors a perpendicular orientation of the moments in the case of Pt-covered films  
contrary to the Co-terminated films with thicknesses below $X = 4$.

We believe that a further thickness increase of the films would not modify the surface contributions to the total MAE. 
This was already seen by Zhang et al.~\cite{Zhang2009} for the Co-terminated case, and we expect the same to occur for the 
Pt-terminated case (see Fig.~\ref{MAE_total}). It is thus interesting to analyze the electronic and 
magnetic structure of the 9-layers thin film, which would be representative of a surface of CoPt alloy. 
In Table ~\ref{table_perfect}, we provide for respectively Co- and Pt-terminated film, the atomic magnetic moments 
and the 
ratio of the change in the interlayer distance $\Delta d$ with respect to the unrelaxed interlayer distance $d_0$. As expected, 
the surface magnetic moments are slightly enhanced in comparison to those characterizing the bulk phase. Moreover, for the considered 
thickness, the bulk moments are recovered in the middle of the films.
In Fig.\ref{DOS_total}, 
we show the atom-projected electronic densities of states (DOS) for the {\bf{L1}$_{0}$} bulk phase and for the surfaces 
with different types of terminations. 
As expected, the Pt band is in general larger than the Co band.  In Fig.\ref{DOS_total}(b), corresponding to 
Co-terminated thin film with $X = 4$, we observe a shrinking of the electronic band width of the outermost Co layer, labeled Co (S), 
because of the  hybridization lowering due to the reduced coordination, which explains the increase of the corresponding magnetic moment. 
However, for the other planes, we recover basically the bulk electronic properties. A similar behavior is obtained 
for Pt-termination as shown in Fig.\ref{DOS_total}(c). 
On the Pt surface layer, labeled Pt (S), the moment increased negligibly with respect to the bulk value but 
the corresponding DOS around the Fermi energy is 
rather different from the bulk counter-part.

Before analyzing the mechanisms and origins behind the increase of the MAE after Pt-coverage, 
we address in the next subsections two possible surface imperfections: stacking fault defects and alloying at the surface. 
%Table 4. 
\begin{table*}
\begin{tabular}{|p{1.2cm}|p{1.2cm}|p{1.2cm}|}
\hline
	Layer & Atom & M (\(\mu_{B}\))\\\hline
	S & Co & 1.88\\\hline
	S-1 & Co & 1.78\\\hline
	S-2 & Pt & 0.35\\\hline
	S-3 & Co & 1.92\\\hline
	Center & Pt & 0.40\\\hline
\end{tabular}
\begin{tabular}{|p{1.5cm}|}
\hline
	\rule[0cm]{0mm}{-0.3cm}\(\delta\)\emph{d}/\(d_{0}\) (\%)\\\hline
	\rule[0cm]{0mm}{-0.3cm}-24.80  \\\hline
	\rule[0cm]{0mm}{-0.3cm} 2.90 \\\hline
	\rule[0cm]{0mm}{-0.3cm}-2.00 \\\hline
	\rule[0cm]{0mm}{-0.3cm}-0.05 \\\hline
\end{tabular}\hspace{1cm}
\begin{tabular}{|p{1.2cm}|p{1.2cm}|p{1.2cm}|}
\hline
	Layer & Atom & M (\(\mu_{B}\))\\\hline
	S & Pt & 0.12\\\hline
	S-1 & Pt & 0.33\\\hline
	S-2 & Co & 1.96\\\hline
	S-3 & Pt & 0.39\\\hline
	Center & Co & 1.93\\\hline
\end{tabular}
\begin{tabular}{|p{1.5cm}|}
\hline
	\rule[0cm]{0mm}{-0.3cm}\(\delta\)\emph{d}/\(d_{0}\) (\%)\\\hline
	\rule[0cm]{0mm}{-0.3cm}12.50  \\\hline
	\rule[0cm]{0mm}{-0.3cm}-3.50 \\\hline
	\rule[0cm]{0mm}{-0.3cm} 0.70 \\\hline
	\rule[0cm]{0mm}{-0.3cm}-0.5 \\\hline
\end{tabular}\\
\caption{Magnetic profile and geometrical relaxations of 
9 layers thick thin films of CoPt  are shown in the left table for the Co-stacking fault and in the right table for the Pt-stacking fault.}
\label{table_stacking}
\end{table*}

\subsection{Stacking faults}

In this section, we consider 9-layers thick thin films with a stacking fault by substituting 
the last atomic layer of the perfect surfaces by the other type of atomic layer. Thus, 
the two possible types of stacking faults would lead to a termination made of Pt 
(Pt-stacking fault) or Co (Co-stacking fault) for the last two surface layers (see Figs.\ref{cell_surf}(c-d)).  
After geometrical relaxation, the interlayer distance at the surface of the Pt-stacking 
fault increases by +12.5 \%, i.e. outward relaxation, 
when compared to the bulk value, unlike the Co-stacking fault, where the relaxation of the surface layer is strongly inward and reaches - 25\% 
(see Table~\ref{table_stacking}). This is undoubtedly due to the fact that Pt atoms are heavier and larger compared to Co atoms. 

In order to compare the MAE obtained in this case to those extracted in the perfect thin films, we consider again the same common reference, i.e. the number of bulk
(CoPt) sequences. The studied Co-stacking fault is then characterized by $X = 3$ (black square in Fig.\ref{MAE_total}) 
while for the Pt-stacking fault $X = 2$ (magenta square in Figs.\ref{MAE_total}).  
One sees that Co-stacking fault leads to a total MAE and a surface contribution, $K_s$, rather similar to the one found 
for a perfect thin film terminated by Co.  This suggests once more the importance of 
having Pt on the surface in order to enhance the MAE. The MAE related to the Pt-stacking fault is larger than the one of a pure 5-layers film terminated by Co but smaller than 
the one of Pt-terminated film. This indicates that increasing the thickness of Pt covering the CoPt films is not necessarily increasing the total MAE. 
Here the surface contribution to 
the MAE is positive and rather large contrary to the case of a Co-stacking fault.

As shown in Table~\ref{table_stacking} (left table), 
the surface magnetic moments in case of Co stacking fault did not increase on the surface unlike the 
perfect thin films. Here, the effect of coordination lowering 
that increases the magnitude of the moment is compensated by the effect of the large inward 
relaxations that favors hybridization of the electronic states and therefore decrease the magnitude of the moment. For the case 
of Pt stacking fault, the results are shown in Table\ref{table_stacking} (right table). Here, the magnetic 
moment of Pt at the outermost surface layer decreases even compared to the bulk one. This is the signature of the induced nature of the Pt moment. 
The closest Co layer to the Pt surface layer is two interlayer distances away and certainly the outward surface relaxation is not a helping factor.

In Figs.~\ref{DOS_total}(d-e), we show the DOS for these surfaces. In Fig.~\ref{DOS_total} (d) corresponding to a Co stacking fault, 
we note that for the Co layer underneath the surface, labeled Co (S-1), the width of the density of states curve is slightly larger than that 
of the bulk. Also, one notices that the {majority}-spin DOS is less occupied for Co (S-1) 
than those of the rest of Co layers. All of that leads to a decrease in the magnetic moment of the S-1 layer. The moment at the surface 
is also not that large compared to the bulk value 
since the interlayer hybridization of the electronic states is rather strong in this particular case. Indeed, 
the surface layer experiences a large inwards relaxation induced by the fact that the bulk lattice 
parameter of Co is much smaller than that of CoPt alloy. We also note that the electronic states of the Pt layers do not 
change dramatically from those of the bulk phase. 
This is certainly not the case in the thin film with Pt stacking fault. 
The Pt surface atom is characterized by a narrower d-band with a larger DOS than that of the bulk as shown in Fig.~\ref{DOS_total} 5(e).  
This is induced by the reduced coordination at the surface. However, the exchange splitting between the majority-- and minority--spin bands  
is smaller than that of 
the bulk. 
This indicates once more the induced nature of the Pt magnetic moment since the closest neighboring Co layer to the 
Pt surface layer is a second nearest neighbor.
%Table 7.
\begin{table*}
\begin{tabular}{|p{1.2cm}|p{2.4cm}|p{2.4cm}|}
\hline
	Layer & Atom & M (\(\mu_{B}\))\\\hline
        S& Pt, Co1, Co2 & 0.33, 2.02, 2.04 \\\hline
        S-1 & Pt & 0.39 \\\hline
        S-2 & Co & 1.93 \\\hline
        S-3 & Pt & 0.40 \\\hline
        Center & Co & 1.93 \\\hline
\end{tabular}
\hspace{1cm}
\begin{tabular}{|p{1.2cm}|p{2.4cm}|p{2.4cm}|}
\hline
	Layer & Atom & M (\(\mu_{B}\))\\\hline
        S& Co, Pt1, Pt2 & 2.00, 0.38, 0.42 \\\hline
        S-1 & Co & 1.95 \\\hline
        S-2 & Pt & 0.39 \\\hline
        S-3 & Co & 1.92 \\\hline
        Center & Pt & 0.40 \\\hline
\end{tabular}
\caption{
Magnetic profile  of thin films considering anti-sites at the surface layer. The results related to the case of Pt anti-site, 
wherein 1/4 of Co surface atoms is replaced by Pt, is shown in the left table and the case of Co anti-site is shown in the right table.
}
\label{table_antisite}
\end{table*}

\subsection{Anti-site defects}

To realize anti-site defects on the surface of the 9-layers thick CoPt thin film, the perfect surface layer was replaced by an alloy of Co$_x$Pt$_{1-x}$, 
with $x = {1}/{4}$ if the originally perfect film is terminated by Pt or $x= {3}/{4}$ for the perfect Co-terminated film. In this case, 
every layer contains 4 atoms in the unitcell as depicted in Figs.\ref{cell_surf}(e-f) and the films are fully relaxed. 
Similarly to the previously studied films, 
here we encounter two 
possibilities: either the sub-surface layer (S-1) is of Pt type or of Co type. The former corresponds then to a Pt anti-site defect while the 
latter is a Co anti-site defect.

The anti-site defects have a dramatic impact on the MAE of the thin films as depicted in Fig.\ref{MAE_total}. Obviously the surface contribution favors 
 strongly the in-plane orientation of the magnetic moments. This is the largest in-plane contribution to the MAE found for all 
investigated films. 

In the case of Pt anti-site, the Pt defect is repelled from the ideal surface position (+6.69 \%)) in line with the result found for Pt stacking fault. 
Pt is a large atom and requires more space, 
which explains this type of 
geometrical relaxation. Note that the Co atoms surrounding the Pt anti-site relax towards the surface (-4.01 \%, -1.66 \%), which leads to a sort 
of surface roughness 
with large deviations in the 
surface atom positions. Unlike Pt anti-site, Co anti-site literally sinks (-14.33 \%) similarly to the surrounding 
neighboring Pt atoms (-11.44 \%, -8.13 \%). The relaxations trends are rather similar to those obtained for the stacking faults. 
 Table~\ref{table_antisite} collects the calculated magnetic moments. Interestingly, the in-plane 
reconstruction was effective for only the subsurface layers in order to accommodate the large surface relaxations. In the case of the 
thin film with Pt anti-site, a slight 
dilatation of 0.74 \% 
is found in contrast to the thin film with Co anti-site where a contraction is noticed (-4.43 \%).
\begin{figure}[ht]
%\centering
 %\begin{tabular}{c}
\includegraphics[width=1.0\columnwidth]{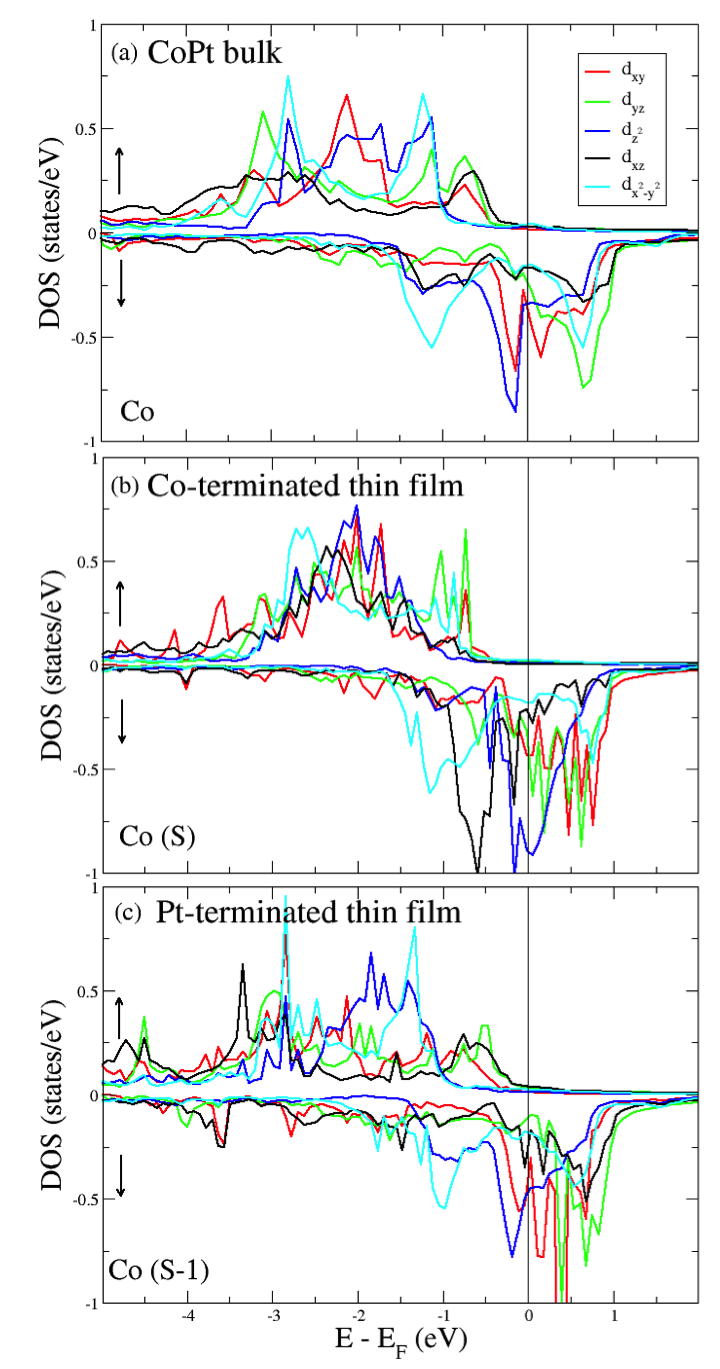}%\hspace{-.8cm}
%\includegraphics[width=0.74\columnwidth]{dos_perfect_Co.png}%\hspace{-.8cm}
%\includegraphics[width=0.74\columnwidth]{dos_perfect_Pt.png} 
%\end{tabular}
\caption{Spin-dependent and orbital resolved density of states of Co in bulk CoPt (a), 
in the outermost perfect surface of CoPt thin film ($X = 3$) shown in (b) and the layer underneath  
the surface layer of the Pt-terminated thin film ($X = 3$) shown in (c).}
\label{DOS_orbital}
\end{figure}

\subsection{Discussion}

Here we discuss the general trend of the MAE for the different investigated systems. The details of the electronic structure 
definitely impact on the SOC related properties. Our aim is to simplify this complicated picture and grasp the main 
ingredients needed to affect the MAE. First we recall that within  the framework of second-order perturbation theory~\cite{Wang1993}:
\begin{equation}
\mathrm{MAE} = \xi^2 \sum_{o,u,\sigma,\sigma'} (2\delta_{\sigma\sigma'}-1)\frac{|\bra{o^{\sigma}}l_z\ket{u^{\sigma'}}|^{2} - |\bra{o^{\sigma}}l_x\ket{u^{\sigma'}}|^{2}}{\epsilon_u^{\sigma}-\epsilon_o^{\sigma'}},
\label{MAE_perturbation}
\end{equation}
where $u^{\sigma}(o^{\sigma'})$ and $\epsilon_u^{\sigma}(\epsilon_o^{\sigma'})$ 
represent eigenstates and eigenvalues of unoccupied (occupied) 
states in spin state $\sigma(\sigma')$; $\xi$ is the SOC constant. $l_z$ and $l_x$ 
are the angular momentum operators. In an alloy like CoPt, 
one has a contribution from each element, Co and Pt, to the MAE given by the 
previous equation. One has to keep in mind that the 
SOC constant $\xi_{\mathrm{Pt}}$ is larger by one order of magnitude than the one 
of Co, $\xi_{\mathrm{Co}}$. The positive and negative contributions 
to the MAE are characterized by $l_z$ and $l_x$ operators, respectively. The 
possible nonzero matrix elements with the $d$-states 
are $\bra{xz}l_z\ket{yz} = 1$, 
$\bra{xy}l_z\ket{x^2-y^2} = 2$, $\bra{z^2}l_x\ket{xz,yz} = \sqrt{3}$, $\bra{xy}l_x\ket{xz,yz} = 1$, 
and $\bra{x^2-y^2}l_x\ket{xz,yz} = 1$. Considering that all majority-spin states 
are occupied, Eq.\ref{MAE_perturbation} is left with two terms only, the one 
involving the coupling between 
the unoccupied and occupied minority-spin states, $(\sigma\sigma') = (\downarrow\downarrow)$, 
and the one involving the coupling between the occupied majority-spin states to 
the unoccupied minority-spin states, $(\sigma\sigma') = (\uparrow\downarrow)$. Interestingly, for the 
$\downarrow\downarrow$($\uparrow\downarrow$)--term 
the $l_z$($l_x$) matrix elements favor an 
out-of-plane easy axis and compete against the $l_x$($l_z$) matrix elements. 
An interesting analysis related to our work is given in the 
context of FeRh films for example.~\cite{Odkhuu2016}

In our discussion we focus on the Co electronic states although {as it will be discussed later on} 
Pt has also a tremendous impact on the 
final MAE. 
We start by analyzing the orbital-resolved DOS for the $d$-states of the Co atoms in CoPt bulk as plotted in 
Fig.~\ref{DOS_orbital}(a). 
One notices that around the Fermi energy, there are minority-spin virtual bound states (VBSs) of large amplitude: $z^2$- 
and $xy$-VBSs 
as well as unoccupied $xy$, $x^2-y^2$ and $yz$. 
With this configuration the matrix elements active and probably important in 
Eq.\ref{MAE_perturbation} for the $\downarrow\downarrow$--term 
are
\begin{equation}
% \frac{|\bra{xz}l_z\ket{yz}|^2}{\epsilon_{yz}-\epsilon_{xz}} 
\frac{|\bra{xy}l_z\ket{x^2-y^2}|^2}{\epsilon_{x^2-y^2}-\epsilon_{xy}}
-  \frac{|\bra{z^2}l_x\ket{yz}|^2}{\epsilon_{yz}-\epsilon_{z^2}}
-  \frac{|\bra{xy}l_x\ket{yz}|^2}{\epsilon_{yz}-\epsilon_{xy}}.
\end{equation}  
{These three terms counteract each other and therefore the  $\downarrow\downarrow$--contribution to the 
MAE is expected to be negligible.} 
  Having a rather localized occupied $z^2$-VBS favors 
an in-plane orientation of the moment, while the occupied $xy$-VBS pushes for an out-of-plane easy axis. 
Considering the $\uparrow\downarrow$--term, we 
expect contributions coming from the majority-spin VBSs: $xy$, $yz$ and $xz$. The active matrix element would then be:
\begin{equation}
% \frac{|\bra{xz}l_z\ket{yz}|^2}{\epsilon_{yz}-\epsilon_{xz}} 
\frac{|\bra{yz}l_x\ket{xy}|^2}{\epsilon_{xy}-\epsilon_{yz}} 
+ \frac{|\bra{xz}l_x\ket{xy}|^2}{\epsilon_{xy}-\epsilon_{xz}} ,
\end{equation}   
where we considered the coupling to the closest unoccupied minority-spin VBS, $xy$, the other possible 
states would lead to rather large 
denominators. {These terms favor the out-of-plane orientation of the magnetic moments}, 
which explains the behavior of bulk CoPt.

For the thin film discussion, we proceed to comparisons involving the same thickness reference, $X=3$.
The orbital-resolved DOS for the $d$-states of the Co-atom at the surface of the Co-terminated thin film is shown in Fig.\ref{DOS_orbital}(b). 
By this analysis, we try to explain why the surface contribution to the total MAE is negative and favors an in-plane easy axis.  
The increase 
of the magnetic moment on the surface compared to the bulk value can be grasped from the larger exchange splitting between the bands. 
Interestingly, important changes occur in both spin channels. In the majority-spin channel, the bulk-VBS close to the Fermi energy and 
favoring the out-of-plane easy axis disappear. Thus, we are left with processes contributing to the $\downarrow\downarrow$--term of the MAE. 
The surface allows to better localize the VBSs pointing out-of-plane, i.e. $z^2$-, $xz$- and $yz$-VBSs, 
which are then less subject to hybridization.
In the minority-spin channel, the $z^2$-VBS becomes prominent. 
Obviously, following the upper discussion for the bulk case, the $l_x$ contribution becomes important favoring then an in-plane orientation 
of the magnetic moment. Moreover the $xy$-VBS decreases in intensity, which is not helping the out-of-plane orientation of the magnetic 
moments. However, a new term involving $l_z$, i.e. favoring the out-of-plane orientation of the moment, shows up and 
involves the coupling between the occupied $xz$-VBS and and the unoccupied $yz$-VBS. Overall, the $l_x$ contribution is certainly more 
important than the one from $l_z$ and thus leading to a negative surface contribution of -0.37 meV to the MAE.

If a Pt-layer is deposited on top of the previous film, we noticed a dramatic increase in the MAE with a switch of the sign of the surface 
contribution. This can be understood from 
the orbital resolved DOS of the Co-atom underneath the Pt surface layer. In this configuration, the DOS shown in Fig.~\ref{DOS_orbital}(c) 
resembles 
more the one 
obtained in bulk CoPt but with slight differences. The $z^2$-VBS decreased in intensity, when compared to the bulk counterpart, thereby 
the $l_x$-contribution in the $\downarrow\downarrow$--term favoring an in-plane orientation of the moment decreases. As shown in 
Table~\ref{table_perfect}, there is a large inward-relaxation 
(-5.6\%) of the Pt surface layer while the Co-layer underneath  relaxes upward (+1.6\%). The interlayer distance between Co and Pt at the 
surface is therefore much smaller than the bulk interlayer distance, 
which affects the intensity of the out-of-plane VBSs. Moreover, the VBSs seen in the bulk 
majority-spin channel are recovered upon deposition of the Pt surface layer. That helps to increase the positive contribution 
of the $\uparrow\downarrow$--term to the MAE. 
\begin{figure*}[htpb]
\centering
 \begin{tabular}{c}
\includegraphics[width=1.1\columnwidth]{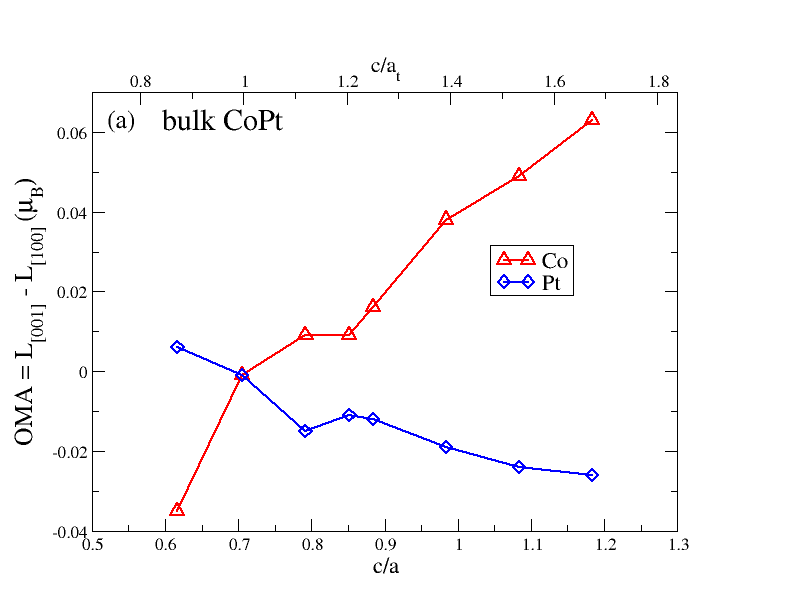}\hspace{-1cm}
\includegraphics[width=1.1\columnwidth]{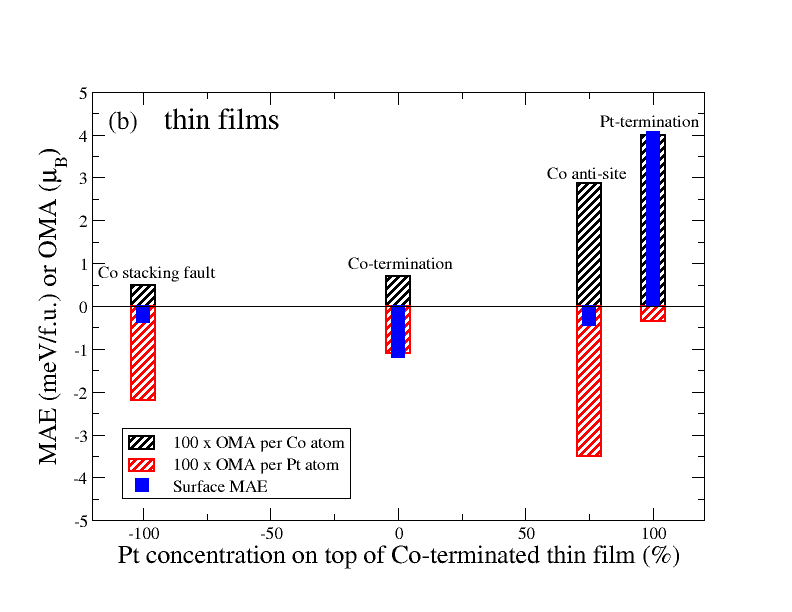}
%\hspace{-.8cm}
\end{tabular}
\caption{(a) Anisotropy of the orbital magnetic moment, $\Delta L = L_{\mathrm{[001]}}- L_{\mathrm{[100]}}$, for Co and Pt 
calculated in the CoPt bulk case. (b) Besides the average Co and Pt OMAs, the surface MAE of the CoPt thin films is plotted as function 
of Pt concentration in the layer covering the Co-terminated thin film with $X = 3$.}
\label{anisotropy_orbital}
\end{figure*}

Another path for the analysis of the calculated MAEs is to use of the celebrated Bruno's formula~\cite{Bruno1989}, which translates to the neglect of 
spin-flip contributions to the MAE as given in Eq.\ref{MAE_perturbation}:
\begin{equation}
\mathrm{MAE} = \sum_i \frac{\xi_i}{4}(L^i_{\mathrm{[001]}} - L^i_{\mathrm{[100]}}),
\end{equation}
where $i$ labels the different atoms, and $L$ being the orbital magnetic moment calculated when the spin magnetic moment points along 
the [100] or the [001] directions. The essence of Bruno's formula is to relate the MAE to the orbital moment anisotropy (OMA), i.e. 
$L_{\mathrm{[001]}} - L_{\mathrm{[100]}}$, and leads to the conclusion that the orientation of the magnetic moments is favored 
when the orbital magnetic moment is maximized. This formula is 
known to work reasonably well when the majority-spin states are occupied. Thus, its validity is 
probably limited to some of the Co atoms discussed in this manuscript but certainly not for Pt. 
It is however instructive to analyze the results obtained with this well 
known formulation since it should correlate with the previous discussion. In Fig.\ref{anisotropy_orbital}(a), the OMAs in the bulk of CoPt 
is plotted for Co and Pt as function of the c/a ratio in fashion similar to that used in Fig.\ref{MAE_bulk}.
One notices that the Pt contribution counteracts the one of Co. While the anisotropy of the Co orbital moment increases with c/a, 
favoring thereby an out-of-plane orientation of the magnetic moment, the anisotropy of 
the Pt orbital moment has an opposite slope and favors an in-plane orientation of the magnetic moment. When summing up the two curves, considering 
the spin-orbit coupling constant, $\xi$, to be the same for Co and Pt, which is of course is not true since $\xi_{\mathrm{Pt}}$ is one order of 
magnitude larger than $\xi_{\mathrm{Co}}$, one recovers the shape of the curve obtained 
in Fig,\ref{MAE_bulk}, i.e. having a minimum of the curve at c/a = 0.8. 

Similar to the bulk, the behavior of the Co and Pt OMA in CoPt thin films counteract each other. In general, the 
Co OMA favors an out-of-plane orientation 
of the moment contrary to the Pt OMA. 
In Fig.\ref{anisotropy_orbital}(b), we plot the surface MAE of the thin films characterized by $X = 3$ as function of the Pt 
concentration on 
the layer deposited on the Co-terminated thin film. Thus, in the case of one perfect Pt overlayer the Pt concentration is 100\%, 
while the 
investigated anti-site corresponds to a Pt concentration of {75\%}. For the specific case of 
Co stacking fault, the Pt concentration is -100\%.
The surface MAE seems to increase  with the Pt concentration but not in a regular manner. 
We plot on the same figure the average Co OMA per 
Co atom and find that this quantity increases smoothly in magnitude with Pt concentration. In addition the contribution of the 
average Pt OMA per Pt atom is shown in  Fig.\ref{anisotropy_orbital}(b). The Pt OMA seems to correlate the irregular 
behavior of the surface MAE. Interestingly, we find that thin films with large Pt OMA per Pt atom 
compared to the Co OMA leads to an in-plane surface MAE. 
Only the Pt-terminated thin film, with a large perpendicular surface MAE, is characterized by a large Co OMA.
 
\section{Conclusion}
We investigated from ab-initio the magnetic behavior of CoPt thin films as function of thickness considering different types of terminations: 
perfect Co or Pt layers or different types of 
defects: anti-site or stacking faults. After this systematic study, we found that the 
MAE is the largest when the thin films are terminated by a perfect Pt 
overlayer. Surprisingly in the latter case, the MAE can be 1000\% times larger than the one of Co-terminated thin films. 
We also find that all types  of investigated defects reduce dramatically the MAE. The surface MAE experiences a sign change when increasing the 
thickness of several investigated films. Except for the Pt-terminated films, the surface MAE favors an in-plane orientation of the moments when the 
thickness $X$ is smaller then four. 
We proceeded to an analysis of the electronic structure of 
the thin films with a careful comparison to the CoPt bulk case and related the behavior of the MAE to the location of the different 
virtual bound states utilizing second order perturbation theory. Finally, the correlation between the MAE and the OMA is studied. 

\section*{Acknowledgments}
We are grateful to Claude Demangeat, Vasile Caciuc, Julen Ibanez-Azpiroz and Manuel dos Santos Dias for helpful discussions. Also we 
thank Hongbin Zhang for discussion and for providing us his data. 
This work was supported by C. N. E. P. R. U project (D 00520090041) of the Algerian government, 
the HGF YIG Program VH-NG-717 (Functional nanoscale 
structure and probe simulation laboratory -- Funsilab) and the ERC Consolidator grant DYNASORE.

\end{document}